\documentclass[11pt]{article}

\usepackage[margin=1in]{geometry}
\usepackage[T1]{fontenc}
\usepackage[utf8]{inputenc}
\usepackage{lmodern}
\usepackage{microtype}
\usepackage{amsmath,amssymb,amsthm}
\usepackage{graphicx}
\usepackage{booktabs}
\usepackage{enumitem}
\usepackage[numbers,sort&compress]{natbib}
\usepackage[hidelinks]{hyperref}
\usepackage{url}
\usepackage{xcolor}

\title{Callability Is Not Operability: Controlled Interface Interventions for LLM Agents}

\author{Zihao Wang \\
Independent Researcher \\
\texttt{zihaow1@seas.upenn.edu}}

\date{July 31, 2026}

\newcommand{\AFT}{\textsc{AFT}}
\newcommand{\AFTBench}{\textsc{AFT-Bench}}

\begin{document}
\maketitle

\begin{abstract} A tool call can be perfectly valid yet still leave an autonomous agent unable to determine what to do next. For example, if an external effect commits but its response is lost, committed and uncommitted states may become indistinguishable to the agent even though they require different continuation actions. We study this gap between \emph{callability} and \emph{operability}: whether a tool interface exposes the action-relevant state and semantics needed for an agent to continue safely under operational uncertainty. We operationalize tool operability through \textbf{Agent-First Tooling (\AFT)}, a set of interface mechanisms spanning selective capability discovery, execution lifecycle and recovery, explicit external-effect semantics, machine-readable results, and postcondition verification. We introduce \textbf{\AFTBench}, a controlled interface-intervention framework that holds the task, backend, initial state, injected failure, agent, and language model fixed while varying the interface exposed to the agent. We first identify mechanism-specific effects in frozen deterministic experiments, replicate key safety findings on a persistent SQLite backend, and then evaluate the same frozen mechanisms with three adaptive LLM families across six workloads. In the pre-specified three-model pooled analysis, selective discovery reduces tool-context exposure by approximately \(4{,}013\) tokens while satisfying a recall non-inferiority criterion. Resumable invocation and durable execution state each improve recovery by \(100\) percentage points under their matched failure treatments. Effect-aware interfaces reduce duplicate effects and unsafe commits by \(56.9\) and \(50.0\) percentage points, respectively, while verification reduces incorrect terminal claims by \(27.8\) percentage points. All seven pooled primary contrasts satisfy their planned inferential criteria after multiplicity correction, although effect magnitudes vary across models. These results show that reliable tool use is not solely a model-capability problem. When operationally distinct states require different actions but are indistinguishable through the interface, stronger model reasoning alone cannot guarantee correct continuation. Tool reliability therefore depends jointly on model policy and the operational semantics exposed at the agent--tool boundary. \end{abstract}

\section{Introduction}

A tool call can be perfectly valid yet still leave an autonomous agent unable to determine what to do next. Consider a side-effecting operation whose request reaches the server and commits successfully, but whose response is lost in transit. From the agent's perspective, the observation may be only a timeout. Yet two underlying states remain possible:
\[
s_{\mathrm{commit}}
\qquad\text{and}\qquad
s_{\mathrm{no\text{-}commit}}.
\]
The safe continuation can differ sharply between them: blindly retrying the first state may duplicate an irreversible effect, whereas refusing to retry the second may leave the task incomplete. If both states produce the same agent-visible observation,
\[
O(s_{\mathrm{commit}})
=
O(s_{\mathrm{no\text{-}commit}}),
\]
then additional model reasoning cannot by itself guarantee the correct continuation action. The missing information lies at the agent--tool boundary.

This example motivates a distinction between \emph{callability} and
\emph{operability}. A tool is \emph{callable} when an agent can identify the
operation and construct a syntactically valid invocation. A tool is
\emph{agent-operable} when its interface exposes enough machine-actionable
state and semantics for an autonomous agent to select an appropriate
capability, execute it over time, recover from failure, reason about external
effects, and verify the resulting state. We use the term
\emph{operational ambiguity} for situations in which different underlying
states require different continuation actions but remain indistinguishable,
or insufficiently actionable, through the interface available to the agent.
Under this view, reliable tool use is not solely a model-capability problem:
it also depends on which action-relevant distinctions the interface preserves
and exposes.

This distinction complements an increasingly rich literature on tool-using
agents. Benchmarks such as BFCL evaluate function selection and invocation,
including stateful multi-turn settings; $\tau$-bench evaluates agents that
interact with users, APIs, policies, and persistent database state; and
ToolSandbox evaluates conversational tool use with implicit state
dependencies and intermediate as well as terminal milestones
\citep{ref13,ref14,ref15}. These benchmarks establish that realistic tool use
is stateful and difficult, but they primarily evaluate the capabilities of
the agent within a given tool environment. Our question is different: when
the task, backend, initial state, failure, controller, and model are held
fixed, how much does changing the \emph{interface semantics themselves}
change what the agent can reliably do?

The tool ecosystem is already evolving toward richer interfaces. On-demand
tool discovery reduces the need to expose entire capability catalogs in the
model context, while programmatic tool execution can reduce repeated
model--tool round trips \citep{ref4,ref5}. The 2026-07-28 revision of the
Model Context Protocol (MCP) further moves the infrastructure boundary:
requests are self-contained under a stateless protocol core, list responses
are cacheable, and the Tasks extension provides durable asynchronous task
handles with explicit status retrieval, update, and cancellation operations
\citep{ref18,ref19}. These developments materially improve connectivity,
scalability, discovery, and long-running execution. They do not by
themselves determine application-specific properties such as whether an
external mutation is idempotent, which preconditions authorize a retry,
whether a stale write should be rejected, or what evidence is sufficient to
establish that a requested postcondition actually holds.

Recent research has begun to isolate such higher-level interface semantics.
Schema-first tool APIs study how formal schemas and structured diagnostics
affect interface adherence under fixed tool semantics \citep{ref16}.
Agent-First Tool API proposes semantic phases for search, resolution,
preview, execution, verification, and recovery \citep{ref9}. Verified Tool
Calls studies postcondition verification, verify-before-retry, and
idempotency under non-atomic failures \citep{ref17}. Atomix and Cordon
introduce transactional mechanisms for containing, staging, committing,
compensating, and auditing effectful agent workflows \citep{ref10,ref11},
while CLI-Anything argues for replacing GUI-centric interaction with
structured, explicit, agent-oriented execution surfaces \citep{ref12}.
Together, these systems indicate that agent reliability increasingly depends
on semantics above basic function invocation. What remains less clear is how
different interface mechanisms contribute under matched conditions, and
whether their effects persist when the same workloads are executed by
adaptive LLM agents rather than deterministic controllers.

We study this question through \textbf{Agent-First Tooling (\AFT)}, an
operational model of the agent--tool interface. Rather than defining a new
transport protocol, \AFT identifies interface mechanisms that expose
action-relevant information across the lifecycle of tool use. The model
includes selective capability discovery, observable execution,
resumable invocation, durable execution state, minimal structured outputs,
explicit side-effect semantics, and postcondition verification. These
mechanisms address different forms of operational ambiguity: which
capability should be used, what happened to an invocation, whether execution
state survives failure, whether an external effect occurred safely, and
whether the agent's terminal belief matches the resulting world state.
Structured outputs provide a machine-readable representation through which
these semantics can be communicated. \AFT is protocol-agnostic: the same
mechanisms can be layered over MCP, conventional function calling, REST
APIs, command-line tools, workflow engines, or adapters around
GUI-oriented systems.

To identify the effect of the interface independently of the model and
backend, we introduce \textbf{\AFTBench}, a controlled interface-intervention
methodology. Within each paired comparison, \AFTBench holds the task,
backend, initial state, injected failure, agent, and language
model fixed while varying the interface exposed to the agent. Interface
conditions form both a cumulative ladder and primitive-level ablations,
allowing mechanism-specific comparisons rather than an aggregate
``agent-friendly'' score. The workloads exercise large capability catalogs,
transient interruption, process-local execution-state loss, response loss
after committed effects, stale state, permission drift, incorrect terminal
reports, and partial completion. Outcomes are evaluated against persistent
world state and include context exposure and capability recall, recovery and
preserved work, duplicate or unsafe external effects, and incorrect terminal
claims. Safe refusal or safe abort is treated as correct when required by the
task policy rather than being collapsed into generic task failure.

Our empirical program separates mechanism identification from ecological
validation. We first freeze deterministic experiments that verify treatment
activation and isolate the corresponding failure mechanisms. We then
replicate key effect-safety results on a persistent SQLite-backed
environment. Finally, without changing the frozen benchmark semantics, we
evaluate adaptive API-backed agents from three contemporary model families
across six workloads. The resulting matrix contains 18/18 completed
model--workload cells and 2,385 result rows.

The results show large, mechanism-specific interface effects. In the
pre-specified three-model pooled analysis, selective discovery reduces
tool-context exposure by approximately \(4{,}013\) tokens while satisfying
the pre-specified recall non-inferiority criterion. Resumable invocation and
durable execution state each improve recovery by \(100\) percentage points
under their matched interruption and state-loss treatments. Effect-aware
interfaces reduce duplicate effects and unsafe commits by \(56.9\) and
\(50.0\) percentage points, respectively, and verification reduces incorrect
terminal claims by \(27.8\) percentage points. All seven pooled primary
contrasts satisfy their planned inferential criteria after multiplicity
correction. The effects are not uniform across models: recovery mechanisms
are stable across all three model families and leave-one-model-out analyses,
whereas the marginal benefit of verification is strongly model-dependent.
This heterogeneity suggests that some interface mechanisms behave like
runtime guarantees, while others partially substitute for model policy.

These results support a more specific claim than the general observation
that ``better tools help agents.'' Tool interfaces determine which
operational distinctions are available to the policy making the next
decision. Stronger models can compensate for some descriptive ambiguity,
but they cannot provide a reliability guarantee when states requiring
different actions remain observationally indistinguishable through the
available interface. Tool-use reliability is therefore jointly determined
by model policy and interface semantics.

\paragraph{Contributions.}
This paper makes three contributions:
\begin{enumerate}[leftmargin=1.5em]

\item \textbf{A formulation of tool operability.}
We distinguish syntactic callability from autonomous operability and
formulate the latter in terms of the action-relevant state and semantics
available to an agent under operational uncertainty. \AFT organizes these
requirements across capability discovery, execution state and recovery,
external-effect semantics, machine-readable results, and verification.

\item \textbf{A controlled interface-intervention methodology.}
We introduce \AFTBench, which isolates interface effects by holding the
task, backend, initial state, failure, controller, and model fixed while
varying interface semantics. Its paired comparisons, cumulative interface
ladder, primitive-level ablations, calibrated fault injection, and
state-based verification are designed to identify mechanism-specific effects
rather than produce a generic tool-use leaderboard.

\item \textbf{Mechanism-level evidence across deterministic and adaptive
agents.}
Using frozen deterministic experiments, a persistent SQLite replication,
and three adaptive LLM families, we find substantial interface effects on
context exposure, interruption and state-loss recovery, duplicate and unsafe
external effects, and terminal-state correctness. Model-stratified and
leave-one-model-out analyses further show that some effects are stable across
models while others exhibit meaningful model--interface interaction.

\end{enumerate}

\section{Related Work}

\paragraph{Agent efficiency beyond task success.} A growing line of work argues that successful task completion alone is insufficient for evaluating practical agents. OSWorld-Human studies the temporal efficiency of computer-use agents and finds that model calls for planning, reflection, and judging account for much of end-to-end latency, while even the strongest evaluated agents require substantially more steps than human-authored trajectories \citep{ref1}. These results highlight inefficiency in agent-side planning and control. We study a complementary source of cost and failure: the interface through which an agent acts. An interface may expose an unnecessarily large capability surface, force additional model--tool interaction, or fail to expose execution state needed for recovery. We therefore treat context exposure and recovery behavior as interface-level variables rather than attributing all inefficiency to the agent policy itself. 

\paragraph{Deployment practice for scalable tool interfaces.} Deployment platforms increasingly incorporate mechanisms for controlling the context and interaction costs of large tool ecosystems. OpenAI recommends reducing sequential request count and parallelizing independent work to reduce round-trip latency, while its MCP integration allows clients to restrict the capability surface imported from a server through \texttt{allowed\_tools}; the documentation notes that exposing many tools can increase both cost and latency \citep{ref2,ref3}. Anthropic similarly provides on-demand tool search rather than requiring complete tool catalogs to be loaded into the model context. Its documentation reports that a representative multi-server configuration can consume roughly \(55{,}000\) tokens in tool definitions before execution begins, and that tool selection quality degrades as the number of simultaneously visible tools grows \citep{ref4}. Anthropic's programmatic tool-calling mechanism addresses a related source of overhead by allowing multiple tool invocations and intermediate transformations to execute within a code-execution loop, without resampling the model or loading every intermediate result into its context \citep{ref5}. These deployment mechanisms provide practical evidence that capability exposure and interaction structure are first-class systems variables for LLM agents. We treat them as motivation rather than as complete operability mechanisms: they primarily address context pressure and interaction overhead, whereas our controlled interventions additionally isolate execution recovery, durable state, external-effect semantics, and postcondition verification under matched failures.

\section{Agent-First Tooling Model}
\label{sec:model}

\subsection{System Model}

We study an LLM-driven agent that acts on an external, stateful environment
through a tool interface. Let the backend environment be

\begin{equation}
    \mathcal{E}
    =
    \left(
        \mathcal{S},
        \mathcal{U},
        P,
        \mathcal{Z}_{E}
    \right),
\end{equation}

where $\mathcal{S}$ is the set of backend states, $\mathcal{U}$ is the set of
backend operations, $P(s' \mid s,u)$ is the transition model induced by
operation $u$, and $\mathcal{Z}_{E}$ is the evidence that can in principle be
obtained from the backend or runtime. Operations may be read-only, reversible,
compensatable, or irreversible.

The backend state is not necessarily exposed directly to the agent. Instead,
the interface $\mathcal{I}$ determines what is observable and which
continuation operations are available. Let $h_t$ denote the execution history
up to step $t$. The interface induces the agent-visible observation

\begin{equation}
    z_t
    =
    \Gamma_{\mathcal{I}}(h_t),
    \qquad
    z_t \in \mathcal{Z}_{\mathcal{I}},
\end{equation}

where $\Gamma_{\mathcal{I}}$ summarizes the information exposed through
schemas, results, lifecycle state, effect metadata, errors, recovery handles,
and verification evidence. Two interfaces over the same backend may therefore
make different distinctions visible to the same agent.

A task is represented as

\begin{equation}
    \tau
    =
    \left(
        x_{\tau},
        s_0,
        \phi_{\tau},
        \alpha_{\tau},
        \psi_{\tau},
        B_{\tau}
    \right),
\end{equation}

where $x_{\tau}$ is the natural-language task specification, $s_0$ the initial
backend state, $\phi_{\tau}(s)$ the requested world-state postcondition,
$\alpha_{\tau}(s,\xi)$ a task-policy acceptance predicate,
$\psi_{\tau}(\xi)$ a trace-level safety predicate, and $B_{\tau}$ the
execution budget.

The distinction between $\phi_{\tau}$ and $\alpha_{\tau}$ allows correct
non-action to be represented explicitly. For example, ``update only if the
record has not changed'' may correctly terminate with a safe abort after a
version conflict, while an action whose authorization has been revoked may
correctly terminate with refusal or escalation rather than mutation.

Given a language model $M$, controller $C$, interface $\mathcal{I}$, backend
$\mathcal{E}$, task $\tau$, and fault schedule $\omega \in \Omega$, execution
produces a trace

\begin{equation}
    \xi
    =
    \operatorname{Run}
    \left(
        M,C,\mathcal{I},\mathcal{E},\tau,\omega
    \right).
\end{equation}

The trace records discovery, model decisions, tool invocations, lifecycle
transitions, recovery attempts, external effects, verification evidence, and
backend-state changes. Faults may include transport interruption, loss of
process-local execution state, response loss after commit, stale state,
authorization changes, incorrect terminal reports, or partial completion.

We separately measure whether the requested state was reached,

\begin{equation}
    Y_{\mathrm{goal}}(\xi,\tau)
    =
    \mathbf{1}
    \left[
        \phi_{\tau}(s_T)=1
    \right],
\end{equation}

and whether execution was policy-correct,

\begin{equation}
    Y_{\mathrm{correct}}(\xi,\tau)
    =
    \mathbf{1}
    \left[
        \alpha_{\tau}(s_T,\xi)=1
        \land
        \psi_{\tau}(\xi)=1
    \right],
\end{equation}

where $s_T$ is the terminal backend state. A run can therefore be correct
because it safely aborts or refuses, while duplicate, unauthorized, or
unintended effects still invalidate an execution.

\subsection{From Callability to Operability}

A tool is \emph{callable} when an agent can identify an operation, construct a
schema-valid invocation, and exchange a protocol-valid request and response.
Callability does not determine what the agent can safely infer or do once
execution becomes uncertain.

Consider two execution histories $h$ and $h'$ that are distinct in the
external world but produce the same observation through interface
$\mathcal{I}$:

\begin{equation}
    \Gamma_{\mathcal{I}}(h)
    =
    \Gamma_{\mathcal{I}}(h').
\end{equation}

Let $\mathcal{A}^{\mathrm{safe}}_{\tau}(h)$ denote the continuation actions
consistent with the task policy and safety requirements after history $h$. We
say that the interface induces \emph{operational ambiguity} when

\begin{equation}
\label{eq:operational-ambiguity}
\begin{aligned}
    \Gamma_{\mathcal{I}}(h)
    &=
    \Gamma_{\mathcal{I}}(h'), \\
    \mathcal{A}^{\mathrm{safe}}_{\tau}(h)
    &\cap
    \mathcal{A}^{\mathrm{safe}}_{\tau}(h')
    =
    \varnothing.
\end{aligned}
\end{equation}

Two underlying situations then require different safe continuations but are
indistinguishable through the available interface. A stronger model may use
prior knowledge to guess which state is more likely, but a policy that receives
the same interface observation cannot guarantee the correct continuation in
both histories.

An interface can reduce this ambiguity in two complementary ways. It can
\emph{distinguish state} by exposing additional evidence---for example,
authoritative lifecycle status, durable invocation state, or postcondition
verification---so that previously indistinguishable histories produce
different observations:

\begin{equation}
    \Gamma_{\mathcal{I}'}(h)
    \neq
    \Gamma_{\mathcal{I}'}(h').
\end{equation}

Alternatively, it can \emph{stabilize continuation}: idempotency, guarded
writes, or related effect semantics can make the same continuation safe across
multiple possible hidden states without revealing which state actually holds,

\begin{equation}
    \mathcal{A}^{\mathrm{safe}}_{\tau,\mathcal{I}'}(h)
    \cap
    \mathcal{A}^{\mathrm{safe}}_{\tau,\mathcal{I}'}(h')
    \neq
    \varnothing.
\end{equation}

Verification primarily improves state distinction; idempotency and guarded
effect semantics primarily stabilize continuation; lifecycle and recovery
mechanisms may do both.

We call an interface \emph{agent-operable} to the extent that it exposes or
enforces enough machine-actionable semantics for an autonomous agent to
discover relevant capabilities, track and continue execution, recover lost
state, act safely on external effects, and verify terminal outcomes.

Operability is not a binary protocol label. We represent it as a
multidimensional profile over an evaluation distribution $q$,

\begin{equation}
    \mathbf{O}(\mathcal{I};q)
    =
    \left(
        R_{\mathrm{capability}},
        P_{\mathrm{recover}},
        S_{\mathrm{effect}},
        V_{\mathrm{outcome}},
        -C_{\mathrm{context}}
    \right),
\end{equation}

where the coordinates summarize capability recall, execution recovery,
effect safety, terminal-outcome correctness, and context cost. Additional
dimensions such as model turns, latency, or human intervention can be appended
when relevant. We retain a vector rather than a single score because different
mechanisms address different failure modes and deployments may value the
coordinates differently.

\subsection{Operational Interface Semantics}

We represent an agent-facing interface as

\begin{equation}
    \mathcal{I}
    =
    \left(
        \mathcal{D},
        \Sigma_{\mathrm{in}},
        \Sigma_{\mathrm{out}},
        \mathcal{L},
        \mathcal{O},
        \mathcal{F},
        \mathcal{R},
        \mathcal{V},
        \mathcal{G}
    \right).
\end{equation}

The components fall into three groups. Capability and representation semantics
$(\mathcal{D},\Sigma_{\mathrm{in}},\Sigma_{\mathrm{out}})$ determine what
operations are exposed and how requests and results are represented. Execution
semantics $(\mathcal{L},\mathcal{O},\mathcal{R})$ define invocation identity,
lifecycle observability, continuation, cancellation, and recovery. Effect and
governance semantics $(\mathcal{F},\mathcal{V},\mathcal{G})$ describe
externally visible effects, idempotency and compensation, verification
evidence, authority, approval, and audit requirements.

This contract is defined above the transport boundary. A transport or
interoperability protocol may encode important portions of the tuple without
determining the full operational semantics of an application. For example,
current MCP specifications provide a stateless protocol core, capability
discovery, structured schemas, and an extension for durable asynchronous task
handling \citep{ref18,ref19}. These mechanisms provide substantial execution
substrate, but they do not generally determine application-specific properties
such as the idempotency scope of an external write, whether a stale mutation is
authorized, whether retry repeats a logical effect, or what evidence
establishes that a requested postcondition holds. \AFT focuses on these
operational semantics rather than replacing the transport protocol.

\begin{table*}[t]
\centering
\small
\renewcommand{\arraystretch}{1.15}
\begin{tabular}{p{3.0cm}p{4.0cm}p{5.1cm}p{3.0cm}}
\toprule
\textbf{Mechanism} &
\textbf{Operational role} &
\textbf{Key interface semantics} &
\textbf{Primary ambiguity} \\
\midrule
Selective discovery &
Find relevant capabilities without exposing the full catalog &
Compact capability metadata, scoped schema materialization, stable identifiers &
Capability selection \\
\midrule
Resumable invocation &
Continue an existing logical operation after interruption &
Stable invocation identity, status retrieval, continuation without re-execution &
Execution continuity \\
\midrule
Observable execution &
Expose authoritative execution progress &
Queryable lifecycle state, ordering/version information, event recovery &
Execution state \\
\midrule
Structured outputs &
Carry operational semantics in machine-readable form &
Status, object references, effect summaries, continuation and verification metadata &
Cross-cutting representation \\
\midrule
Effect semantics &
Make external continuation safe under uncertain effects &
Idempotency, preconditions, effect scope, commit point, authority, compensation &
Effect safety \\
\midrule
Recoverable state &
Reconstruct invocation state after runtime-state loss &
Durable state, recovery handles, reconstruction and reconciliation semantics &
Execution-state loss \\
\midrule
Postcondition verification &
Check whether terminal claims match resulting world state &
Authoritative evidence, effect references, freshness, partial-outcome evidence &
Outcome correctness \\
\bottomrule
\end{tabular}
\caption{Interface mechanisms in \AFT. The mechanisms address different forms
of operational ambiguity rather than defining a single monotonic notion of
agent-firstness.}
\label{tab:aft-mechanisms}
\end{table*}

The table summarizes the seven mechanisms without implying that they contribute
equally in every workload. Three distinctions are particularly important for
the experiments in this paper.

Resumability and durability are separate properties. Resume specifies how a
client continues a known invocation after interruption; durable execution
state determines whether that invocation can still be reconstructed after
process-local state has been lost.

Effect semantics and verification also solve different parts of post-call
uncertainty. Effect semantics can stabilize continuation without revealing the
hidden state---for example, by making retry idempotent or a write conditional
on a version. Verification instead supplies evidence about which outcome
actually occurred and can correct an earlier success, failure, or unknown
claim.

Structured output is cross-cutting rather than an independent safety
guarantee. It is the machine-readable carrier through which lifecycle, effect,
recovery, and verification semantics are communicated.

\subsection{Mechanisms as Ambiguity Resolution}

The mechanisms above admit a common operational interpretation. Selective
discovery preserves capability-relevant distinctions while reducing context
exposure. Observable execution, resumable invocation, and recoverable state
expose or reconstruct execution-state distinctions. Effect semantics stabilize
actions such as retry and guarded mutation across uncertain effect states.
Postcondition verification restores distinctions between reported and actual
outcomes. Structured outputs provide the representation through which these
semantics are communicated.

This view does not imply that every agent failure is an information problem,
nor that every interface mechanism must reveal hidden state. The common
criterion is narrower: whether the interface gives the agent and its runtime a
reliable way to choose a safe continuation when execution becomes uncertain.

\begin{figure}[t]
    \centering
    \includegraphics[width=\linewidth]{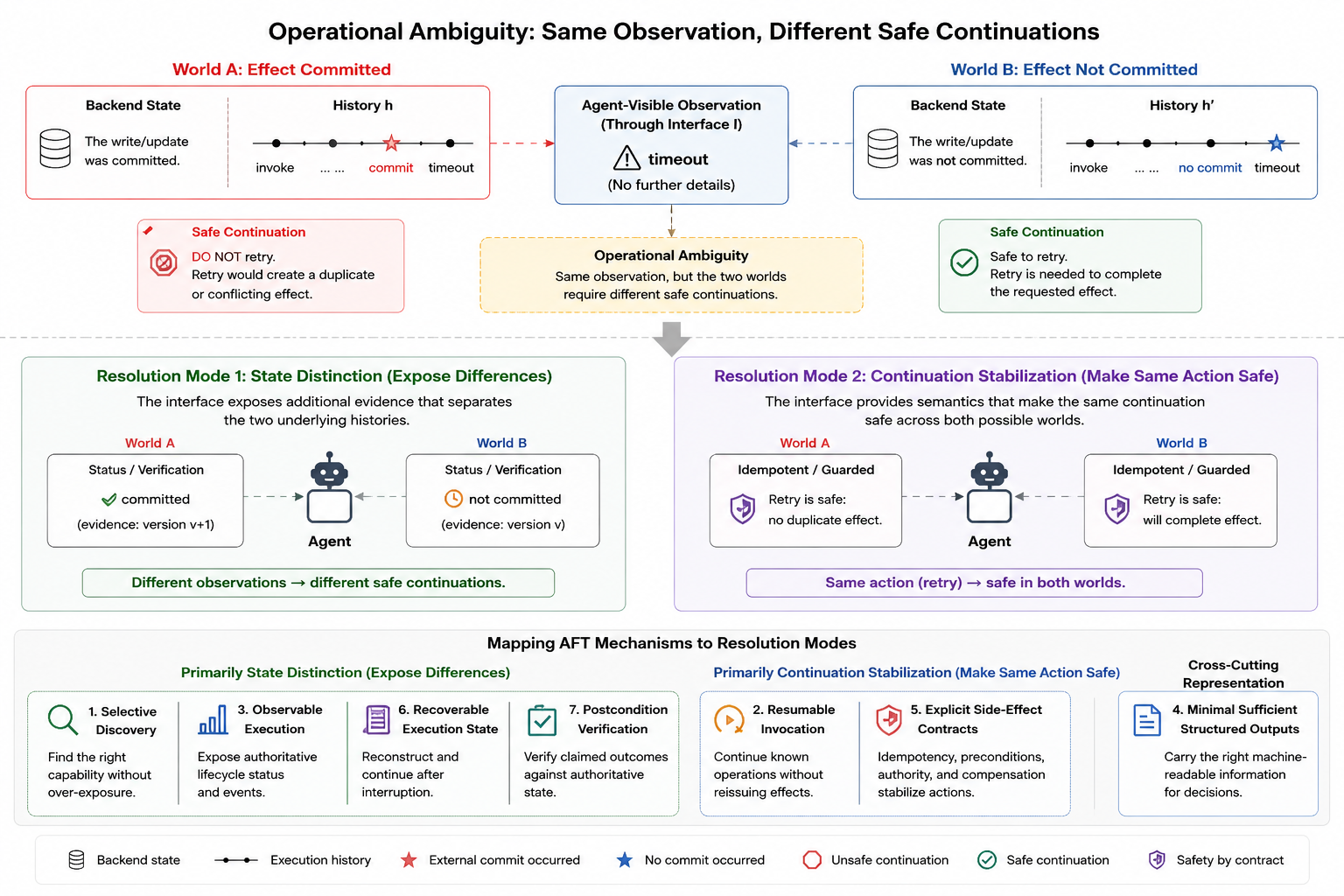}
    \caption{
    Callability does not guarantee operability. Operational ambiguity arises
    when different underlying histories require different safe continuations
    but produce the same interface-visible observation. An interface can
    improve operability either by exposing evidence that distinguishes those
    histories or by providing continuation semantics that remain safe across
    them.
    }
    \label{fig:operational-ambiguity}
\end{figure}

\section{Experimental Design}
\label{sec:experimental-design}

Our empirical design treats the tool interface as the experimental treatment.
The central comparison is deliberately simple: the task, backend state,
failure, controller, model, and execution budget remain fixed, while the
interface semantics exposed to the agent change. This paired design allows
differences in recovery, safety, or context consumption to be attributed to
the interface rather than to task difficulty or model capability.

A run is indexed by

\begin{equation}
    e =
    (w,\tau,\mathcal{I},M,C,\omega,r),
\end{equation}

where $w$ is the benchmark world, $\tau$ the task, $\mathcal{I}$ the interface
condition, $M$ the model when an adaptive LLM is used, $C$ the controller,
$\omega$ the injected fault schedule, and $r$ the repetition index. Paired
runs additionally match the initial state, task manifest, backend version,
fault schedule, controller version, model identifier, and source/configuration
provenance. Only the interface treatment relevant to the comparison is
allowed to differ.

\subsection{Evidence Layers and Experimental Controls}

The evaluation has three layers. First, a deterministic capability-aware
controller identifies mechanism effects under controlled faults. This stage
validates fault activation, manipulation checks, recovery paths, safety labels,
and treatment isolation without stochastic model behavior. The resulting
deterministic evidence was formally frozen before any adaptive LLM validation.
After this freeze, the task definitions, interface treatments, fault semantics,
controller contract, verifiers, and primary endpoints were held fixed
throughout the LLM phase.

Second, key effect-safety findings are repeated on a persistent SQLite-backed
environment. This provides a production-like check that duplicate effects,
stale updates, and unsafe commits remain meaningful when state transitions
cross a durable transaction boundary rather than an in-memory simulator.

Third, the frozen benchmark is executed with three adaptive API-backed model
families: Qwen 3.7 Plus, DeepSeek V4 Pro, and GPT-5.6 Sol. Each model is
evaluated on six canonical workloads covering selective discovery,
interruption recovery, post-commit response loss, stale-state and permission
drift, terminal-outcome verification, and partial-success verification. All
\(3\times6=18\) model--workload cells completed, yielding \(2{,}385\) result
rows. No task semantics, interface treatment, fault semantics, verifier, or
endpoint was modified after the evidence freeze.

All benchmark worlds use resettable task-specific initial states and an
append-only oracle trace. Within a pair, the backend implementation and
business operations are identical. Stronger interfaces may expose additional
lifecycle, recovery, effect, or verification semantics, but they may not add a
new business capability or perform the requested task on behalf of the agent.

\subsection{Interface Treatments and Workloads}

The deterministic benchmark includes a cumulative interface ladder
\(I_0\)--\(I_5\) together with mechanism-removal conditions. The ladder is
useful for implementation and manipulation checks, but the main analysis does
not interpret it as a single monotonic ``agent-firstness score.'' Instead, we
compare individual mechanisms under the failure modes they are intended to
address.

The seven primary mechanism comparisons are summarized in
Table~\ref{tab:primary-contrasts}. We use descriptive names throughout the
paper rather than internal hypothesis identifiers.

\begin{table}[t]
\centering
\small
\renewcommand{\arraystretch}{1.18}
\begin{tabular}{p{2.8cm}p{4.0cm}p{5.4cm}}
\toprule
\textbf{Comparison} &
\textbf{Interface treatment} &
\textbf{Workload and primary endpoint} \\
\midrule
Discovery context &
selective vs.\ eager discovery &
large catalog; exposed tool-context tokens \\
\midrule
Discovery recall &
selective vs.\ eager discovery &
large catalog; capability recall, tested for non-inferiority \\
\midrule
Resume recovery &
full vs.\ resume removed &
transient interruption; recovery success \\
\midrule
Durable-state recovery &
full vs.\ durable state removed &
process-local state loss; recovery success \\
\midrule
Duplicate effects &
effect-aware vs.\ legacy interface &
post-commit response loss; duplicate logical effects \\
\midrule
Unsafe commits &
strong vs.\ weak effect semantics &
stale state and permission drift; unsafe committed actions \\
\midrule
Terminal verification &
verification vs.\ verification removed &
incorrect terminal reports; incorrect terminal claims \\
\bottomrule
\end{tabular}
\caption{Primary mechanism comparisons. Each pair preserves the same task,
backend, initial state, controller, model, and fault realization while changing
only the interface semantics relevant to the targeted mechanism.}
\label{tab:primary-contrasts}
\end{table}

The discovery workload scales the tool catalog over
\(|\mathcal{D}| \in \{10,50,200,1000\}\).
Eager conditions expose the complete baseline tool surface, whereas selective
conditions expose compact capability metadata and materialize full schemas
only when required. We record context exposure, schema materialization,
top-\(k\) recall, fallback behavior, and final capability selection. Recall is
measured independently of task completion so that token reduction cannot be
mistaken for successful discovery.

The discovery analysis therefore asks two complementary questions: whether
selective discovery reduces context exposure, and whether that reduction
preserves capability recall. Recall is evaluated using the pre-specified
non-inferiority margin

\begin{equation}
    \delta = 0.10.
\end{equation}

The recovery workloads distinguish two failure classes. A transient
interruption preserves the logical invocation and tests whether execution can
resume without restarting the effect. A process-local state-loss condition
removes runtime-local state while allowing durable execution state, when
supported by the interface, to survive. This separation is important because
resumability and durability solve different operational problems.

Effect-safety workloads cover response loss after a committed effect, stale
state, and permission drift. Verification workloads additionally inject
incorrect terminal reports: the response channel may report success although
the requested postcondition is false, or report failure although the requested
state already holds. A partial-success variant is analyzed separately as a
secondary robustness test rather than as an additional primary comparison.

\subsection{Fault Injection and Outcome Semantics}

Faults are injected by the harness at backend-defined execution boundaries,
not generated through natural-language prompts. The trace records whether the
request was accepted, backend execution began, the external commit point was
crossed, a response was generated and delivered, and subsequent recovery or
reconciliation changed the state.

For post-commit response loss, the canonical sequence is

\begin{equation}
\begin{aligned}
&\texttt{REQUEST\_ACCEPTED}
\rightarrow
\texttt{BACKEND\_STARTED}
\rightarrow
\texttt{EFFECT\_COMMITTED} \\
&\qquad\rightarrow
\texttt{RESPONSE\_GENERATED}
\rightarrow
\texttt{RESPONSE\_DROPPED}.
\end{aligned}
\end{equation}

The benchmark distinguishes transport retransmission, continuation of an
existing invocation, logical re-execution of the task action, and
reconciliation against authoritative state. These operations are recorded
separately because two interfaces may both issue another request after a
timeout while only one preserves the original logical effect.

Outcome labels are derived from persistent backend state and the oracle trace,
not from the agent's final message. A run may terminate as completed,
safely aborted, safely refused, safely escalated, unsafely committed,
unnecessarily failed, or unresolved. Safe non-action counts as correct when
required by the task policy. Safety metrics further identify duplicate,
unauthorized, unintended, or residual effects, while verification metrics
identify false success, false failure, and other incorrect terminal claims.

Before canonical evidence generation, these metrics were calibrated with
known-positive and known-negative cases. This prevents an apparent zero
violation rate from arising simply because a detector never activates.

\subsection{Agent Execution and Instrumentation}

The execution stack separates model decisions from runtime guarantees. The
model policy performs semantic interpretation, capability selection, argument
construction, and high-level continuation decisions. The runtime enforces
budgets, dispatches calls, injects faults, records traces, and applies only
those lifecycle, recovery, idempotency, or verification guarantees exposed by
the active interface condition.

This distinction prevents the harness from silently repairing a weak
interface. For example, the legacy condition does not receive logical-effect
deduplication simply because the runtime is capable of implementing it, and a
verification-removed condition cannot query authoritative postcondition
evidence through a hidden path.

Each run emits an append-only trace containing the model-visible capability
surface, tool invocations, lifecycle events, backend stages and commit points,
dropped responses, retries or resumes, reconciliation and verification events,
external-effect identities, and the terminal backend state. Timing is recorded
with monotonic high-resolution clocks. We use these measurements for local
stage-level accounting only; the study does not claim production-scale
network or queueing latency.

Adaptive LLM runs use the same controller contract within each paired
comparison. Across the completed matrix, \(7{,}524\) provider calls generated
35 retained provider or transport failures. Failed calls produced no synthetic
benchmark rows: incomplete tuples were resumed under an append-only,
fail-closed policy. All 18 cells were ultimately completed with zero fabricated
rows and zero duplicate result tuples.

One discovery experiment reached its initial non-binding monetary safety
ceiling before completion. Under the budget rule specified for the evaluation,
the ceiling was increased and the same unfinished experiment resumed without
changing its model, task, prompt, interface, fault, seed, verifier, treatment,
or endpoint. We therefore treat this event as an operational continuation
rather than a new experimental condition.

\subsection{Statistical Analysis}

The pooled LLM analysis uses the seven mechanism-specific comparisons
summarized in Table~\ref{tab:primary-contrasts}. The analysis plan was fixed
before inspection of the combined three-model results, although it was not
publicly preregistered.

All effects are paired at the task--fault--model level. Metrics are assigned a
utility direction before aggregation so that positive values always favor the
agent-first treatment. Thus recovery and recall are higher-is-better, whereas
context exposure, duplicate effects, unsafe commits, and incorrect terminal
claims are lower-is-better.

For each comparison we report the number of matched pairs, paired utility
difference, a task-clustered \(95\%\) bootstrap interval based on \(2{,}000\)
resamples, and a paired sign-flip permutation test using \(10{,}000\)
permutations. The seven primary comparisons form one multiplicity family and
are corrected with the Holm procedure. Discovery recall is evaluated as a
one-sided non-inferiority comparison with the pre-specified margin
\(\delta=0.10\).

Model-specific effects and model--interface interactions are secondary
analyses. Multiple model-specific comparisons use Benjamini--Hochberg
false-discovery control. Because the interaction analysis contains only eight
task clusters, it is used for interpretation rather than promoted to a
confirmatory claim.

We additionally perform leave-one-model-out analyses over the three pairwise
model subsets to determine whether any pooled conclusion is driven by a single
model family. These analyses show that recovery effects are especially stable
across models, whereas the marginal benefit of verification is more
model-dependent.

\subsection{Evidence Freeze and Reproducibility}

The deterministic evidence was formally frozen before adaptive LLM evaluation.
The complete deterministic and LLM evidence was subsequently placed under an
experimental lock after the planned model matrix, heterogeneity analyses,
provider-error audit, and release checks were completed. No additional API
calls or new experimental families were introduced after this point.

The release contains resolved experiment configurations, task and state
manifests, raw results, operational traces, manipulation checks, statistical
outputs, and source w/ configuration provenance. Legacy results produced before
the measurement-validity audit are preserved for provenance but excluded from
all canonical analyses.

Numerical values reported in the paper are generated automatically from the
structured analysis outputs rather than copied manually into the manuscript.
The final release audit verifies completion and uniqueness of the full
model--workload matrix, consistency between results and manifests, exclusion
of invalidated legacy evidence, absence of fabricated rows following provider
failures, and agreement between structured analysis outputs and manuscript
values.

At experimental lock, the implementation passed 465 automated tests together
with linting, compilation, repository-integrity, and manuscript-consistency
checks. Provider and transport failures were retained in an append-only audit
trail and never converted into synthetic benchmark observations.

\subsection{Threats to Validity}

The study is designed for \emph{mechanism identification}, not for estimating
how frequently each failure occurs in production. The workloads intentionally
activate conditions in which the corresponding interface semantics should
matter, so the reported effects should be interpreted as conditional mechanism
effects rather than real-world failure prevalence.

Synthetic worlds provide strong experimental control but cannot represent the
full heterogeneity of deployed services. The SQLite replication confirms key
safety effects across a persistent transaction boundary, but does not establish
backend-independent universality. Likewise, the adaptive evaluation covers
three contemporary model families rather than all current or future LLMs;
model-specific and leave-one-model-out results are therefore reported
explicitly.

Hosted-model behavior may drift over time despite exact identifier and
provenance recording. The benchmark also measures local structural timing
rather than production-scale network or queueing performance. Finally,
observable execution and minimal structured outputs are part of the broader
\AFT model, but their independent efficacy evidence is weaker than the primary
results for discovery, resume, durable state, effect semantics, and
verification. We therefore treat dedicated event-loss and parser-repair
experiments as future extensions rather than as current headline findings.

\section{Results}
\label{sec:results}

We report the adaptive three-model results as the main ecological validation
of the frozen interface mechanisms, while using the deterministic experiments
to establish treatment activation and mechanism identity. All seven
pre-specified pooled comparisons satisfy their planned inferential criteria
after multiplicity correction. Throughout this section, effect sizes are
reported as \emph{utility differences}: positive values favor the
agent-operable interface treatment, regardless of whether the underlying raw
metric is higher-is-better (for example, recovery) or lower-is-better (for
example, duplicate effects).

Table~\ref{tab:main-results} summarizes the pooled results. The findings are
mechanism-specific rather than evidence of blanket superiority for one
interface condition. Discovery primarily changes context exposure, recovery
semantics dominate interruption and state-loss workloads, effect contracts
change the safety of external actions, and verification changes the accuracy
of terminal claims.

\begin{table*}[t]
\centering
\small
\renewcommand{\arraystretch}{1.15}
\begin{tabular}{p{3.5cm}r p{3.3cm} p{4.1cm} r}
\toprule
\textbf{Comparison} &
\textbf{Pairs} &
\textbf{Utility difference} &
\textbf{95\% task-clustered CI} &
\textbf{Adjusted $p$} \\
\midrule
Discovery context &
72 &
\(4013.17\) fewer tokens &
\([1976.14,\ 6231.28]\) &
0.0007 \\
\midrule
Discovery recall &
72 &
\(0.0139\) &
\([0.0000,\ 0.0417]\) &
0.0007 \\
\midrule
Resume recovery &
72 &
\(1.0000\) &
\([1.0000,\ 1.0000]\) &
0.0007 \\
\midrule
Durable-state recovery &
72 &
\(1.0000\) &
\([1.0000,\ 1.0000]\) &
0.0007 \\
\midrule
Duplicate effects &
72 &
\(0.5694\) &
\([0.3750,\ 0.7639]\) &
0.0007 \\
\midrule
Unsafe commits &
72 &
\(0.5000\) &
\([0.3056,\ 0.6806]\) &
0.0007 \\
\midrule
Terminal verification &
144 &
\(0.2778\) &
\([0.1806,\ 0.3750]\) &
0.0007 \\
\bottomrule
\end{tabular}
\caption{Pooled three-model primary results. Positive utility differences
favor the agent-operable treatment. Discovery recall is evaluated using the
pre-specified one-sided non-inferiority criterion with margin
\(\delta=0.10\); the remaining primary comparisons use paired two-sided
inference. All primary comparisons are adjusted jointly using the Holm
procedure.}
\label{tab:main-results}
\end{table*}

\subsection{Selective Discovery Compresses Context Without Sacrificing Recall}

Selective discovery substantially reduces the amount of tool information that
must be placed in the model context. Across 72 matched pairs, the selective
interface reduces tool-context exposure by \(4013.17\) tokens on average
relative to eager exposure, with a task-clustered \(95\%\) confidence interval
of \([1976.14,6231.28]\) tokens. The effect remains positive after
multiplicity correction.

The reduction in context is not obtained by making the correct tool harder to
find. The paired recall difference is \(0.0139\), with a \(95\%\) confidence
interval of \([0.0000,0.0417]\). The lower bound remains well above the
pre-specified non-inferiority margin of \(-0.10\). We therefore reject the
interpretation that selective discovery merely trades capability recall for
token savings: within the evaluated catalogs, it preserves task-relevant
discovery while sharply reducing the interface surface presented to the
model.

This result is important for the distinction between callability and
operability. Eager exposure and selective discovery can make the same
underlying capabilities callable, yet they impose very different information
costs on the agent. The interface can therefore preserve the distinctions
needed for capability selection without materializing the complete schema
surface in the model context.

\begin{figure*}[t]
    \centering
    \includegraphics[width=0.96\textwidth]{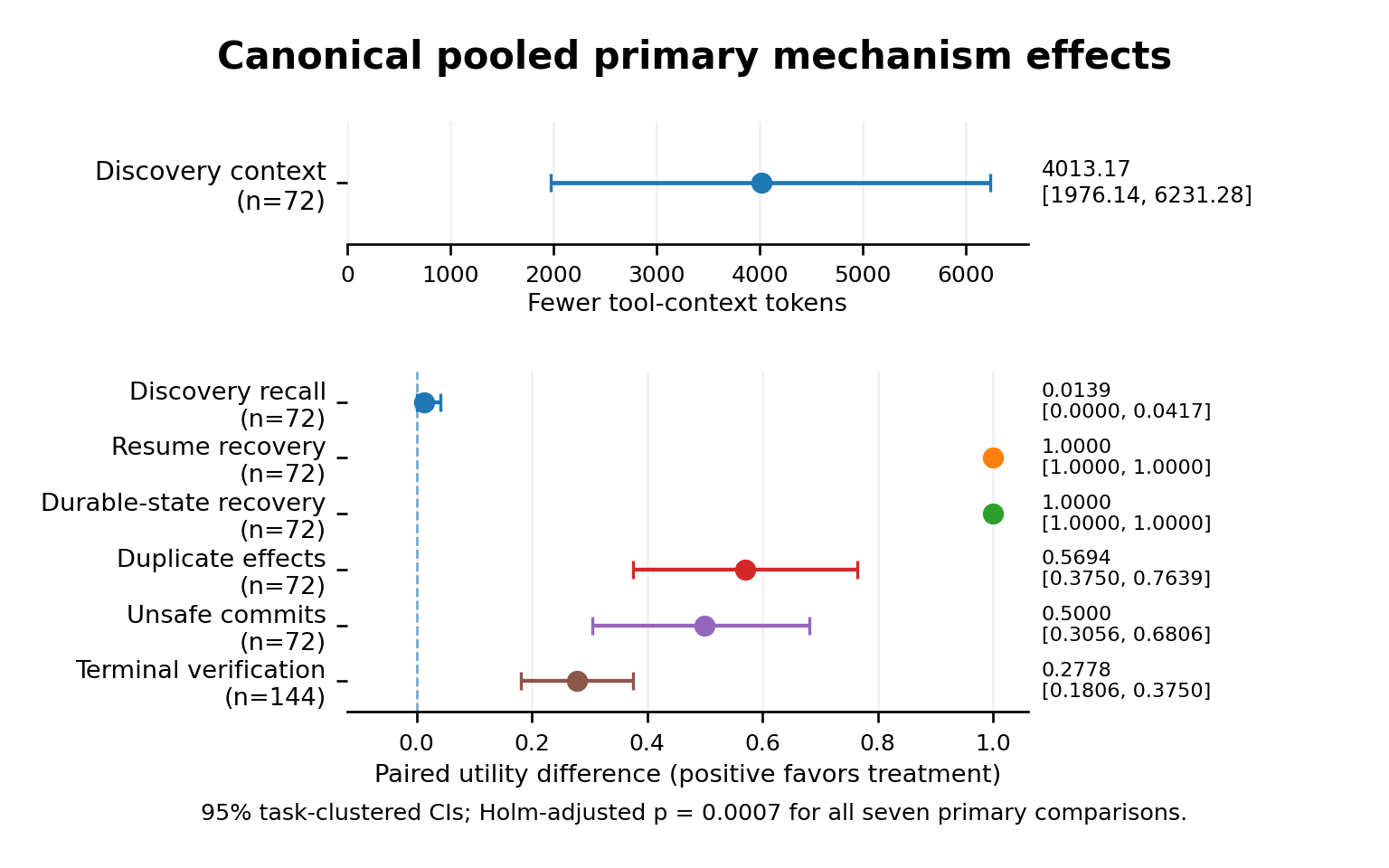}
    \caption{Canonical pooled primary mechanism effects. The discovery-context
    panel reports fewer exposed tool-context tokens; the remaining panels
    report paired utility differences, where positive values favor the
    agent-operable treatment. Error bars show 95\% task-clustered confidence
    intervals. Pair counts, effect estimates, and intervals match
    Table~\ref{tab:main-results}.}
    \label{fig:pooled-primary}
\end{figure*}

Figure~\ref{fig:pooled-primary} summarizes the canonical pooled primary
mechanism effects reported in Table~\ref{tab:main-results}.

\subsection{Recovery Depends on Matching Interface State to Failure Mode}

The recovery experiments distinguish transient interruption from loss of
process-local execution state. The distinction matters because resumability
and durability address different operational failures.

Under transient interruption, removing resumable invocation reduces recovery
by the full \(1.00\) utility unit across all 72 matched pairs. The
task-clustered interval is \([1.00,1.00]\). In other words, the contrast is
saturated in this workload: the resumable condition recovers while the
resume-removed condition does not for every matched pair represented in the
pooled analysis.

Durable execution state exhibits the same magnitude under the state-loss
treatment. Removing durable state reduces recovery by \(1.00\), again with a
task-clustered interval of \([1.00,1.00]\). These effects should not be read
as claims that either mechanism guarantees recovery under arbitrary failures.
Rather, they show that the two mechanisms are decisive under the specific
failure classes they are designed to address.

The separation provides a useful systems interpretation. A transient
connection or controller interruption can leave the logical invocation
intact, in which case a stable invocation identity and resume operation are
sufficient. Once process-local execution state is lost, continuation requires
state that survives outside that process. Treating both cases as generic
``retry failures'' would hide this distinction.

These recovery effects are also the most stable findings across model
families. They appear with the same direction and maximal effect in all three
model-specific analyses and remain unchanged under every leave-one-model-out
combination. This stability suggests that resumability and durable state
behave primarily as runtime properties: stronger model policies do not remove
the need for execution state that the interface or runtime must preserve.

\subsection{Effect Semantics Prevent Duplicate and Unsafe Continuations}

The strongest safety results arise when an external action becomes uncertain
after the agent has already issued a valid call. In the post-commit
response-loss workload, the backend commits the effect before the response is
dropped. A legacy interface therefore exposes an ambiguous timeout without a
strong guarantee about whether re-execution will repeat the logical effect.

Effect-aware interfaces reduce duplicate effects by \(0.5694\) in pooled
utility, corresponding to a \(56.94\)-percentage-point reduction in the
paired duplicate-effect outcome. The \(95\%\) confidence interval is
\([0.3750,0.7639]\). The result shows that explicit effect identity and
idempotency semantics materially change the safety of continuation after a
lost response.

A related pattern appears when state or authority changes between planning
and execution. Stronger effect semantics reduce unsafe commits by \(0.5000\),
with a \(95\%\) confidence interval of \([0.3056,0.6806]\). Here the important
outcome is often \emph{non-action}: guarded execution can turn what would have
been a stale or unauthorized mutation into a safe abort or refusal. Evaluating
only whether the originally requested mutation occurred would therefore
misclassify an important form of correct autonomous behavior.

The persistent SQLite replication shows the same qualitative safety pattern
across a durable transaction boundary. The weaker interface conditions
produce nine duplicate effects and three unsafe commits, whereas the
effect-aware conditions produce none. No implementation errors are observed
in this replication. The SQLite experiment does not establish universal
backend independence, but it reduces the possibility that the safety effects
are artifacts of an in-memory simulator or synthetic commit flag.

Model-specific results reveal more variation here than in the recovery
experiments. The duplicate-effect benefit is smaller for GPT than for the
other two model families, and its model-specific exploratory evidence is
correspondingly weaker. The pooled effect remains positive, but this
heterogeneity cautions against treating effect-contract value as a constant
independent of agent policy.

\begin{figure*}[t]
    \centering
    \includegraphics[width=0.96\textwidth]{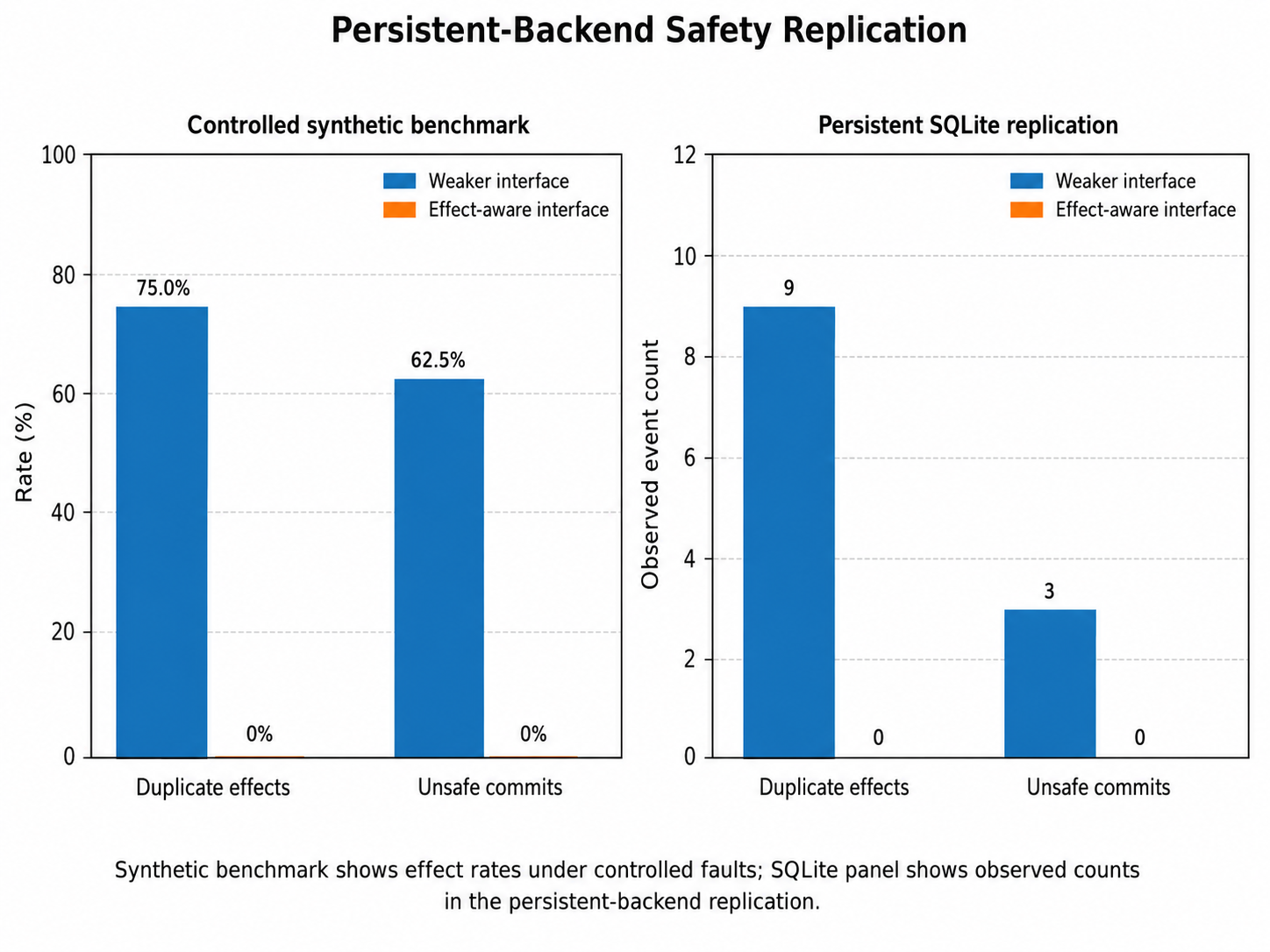}
    \caption{Persistent-backend safety replication. Effect-aware interfaces
    retain the same qualitative advantage for duplicate-effect and
    unsafe-commit outcomes when the experiment is repeated across a durable
    SQLite transaction boundary.}
    \label{fig:sqlite-replication}
\end{figure*}

Figure~\ref{fig:sqlite-replication} compares the controlled synthetic
findings with the persistent-backend replication.

\subsection{Verification Corrects Incorrect Terminal Beliefs}

Execution can complete while the agent's belief about the outcome remains
wrong. The verification workloads explicitly separate the reported terminal
result from the authoritative world state: a response may claim success when
the requested postcondition is false, or report failure or uncertainty when
the requested state already holds.

Postcondition verification reduces incorrect terminal claims by \(0.2778\)
in pooled utility across 144 matched pairs, with a \(95\%\) confidence
interval of \([0.1806,0.3750]\). This result captures a different kind of
recovery from resume or durable state. Resume repairs an interrupted
execution; verification repairs the agent's belief about what the execution
actually accomplished.

A secondary partial-success workload strengthens this interpretation. When a
multi-part effect is only partially realized but the response overstates
completion, verification improves terminal-claim correctness by \(0.5185\)
across 27 pooled pairs, with a \(95\%\) confidence interval of
\([0.2593,0.7778]\). This robustness comparison lies outside the primary Holm
family and is reported with unadjusted two-sided \(p=0.0001\).

Verification is also the clearest example of model-dependent marginal value.
Its utility difference is \(0.3542\) for Qwen and \(0.4792\) for DeepSeek,
but \(0\) for GPT under the evaluated workload. Trace-level diagnostics show
that the GPT condition still receives and can use verification evidence; the
zero marginal effect instead coincides with a more conservative terminal
reporting policy that already avoids the targeted incorrect claims. We
therefore interpret this pattern as evidence of partial substitution between
model policy and explicit verification, rather than as missing verification
instrumentation.

This observation places an important boundary on the broader claim that
interface semantics matter. Better models can reduce the frequency of some
errors and thereby reduce the marginal value of a corresponding interface
mechanism. They do not, however, make all interface mechanisms redundant:
the recovery results above remain unchanged across models, and effect-safety
benefits remain positive in the pooled analysis.

\subsection{Interface Effects Persist Across Models but Are Not Uniform}

\begin{figure*}[t]
    \centering
    \includegraphics[width=0.96\textwidth]{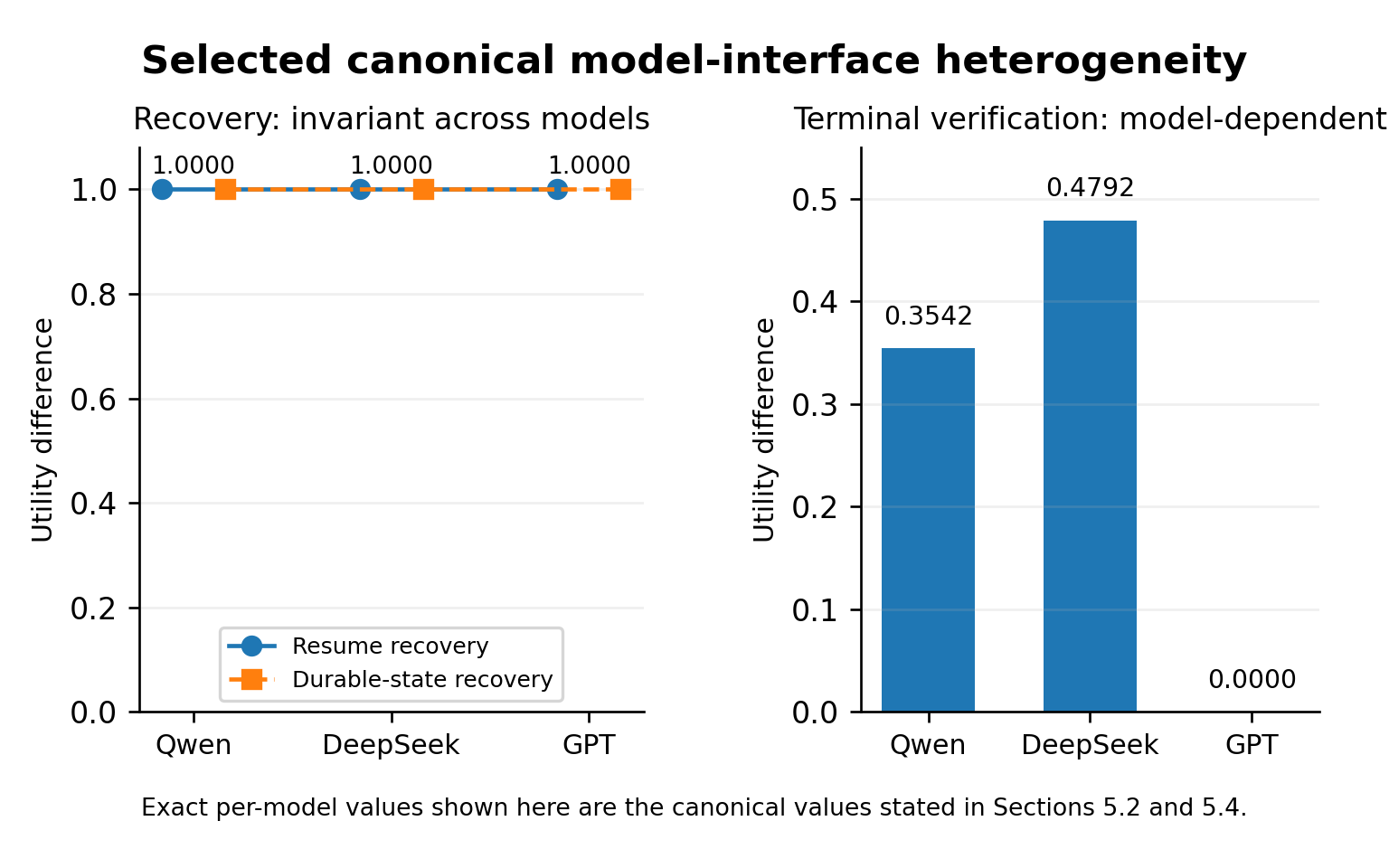}
    \caption{Selected canonical model--interface heterogeneity across the
    three adaptive LLM families. Resumable invocation and durable-state
    recovery have utility difference 1.0000 for each model; terminal
    verification has utility difference 0.3542 for Qwen, 0.4792 for DeepSeek,
    and 0.0000 for GPT, matching the exact per-model values stated in the text.}
    \label{fig:model-heterogeneity}
\end{figure*}

Figure~\ref{fig:model-heterogeneity} highlights the model-specific effects
for which exact canonical per-model values are stated in the text: recovery
and terminal verification. The main pattern is not uniform improvement across
every model and endpoint, but a combination of robust runtime effects and
model-dependent policy effects.

Resumable invocation and durable execution state are invariant across the
three evaluated model families and across all leave-one-model-out analyses.
Discovery effects also preserve the same qualitative context--recall tradeoff.
By contrast, effect safety is weaker for GPT, and terminal verification varies
substantially by model. The interaction analysis is therefore useful for
interpretation but not treated as confirmatory evidence: only eight task
clusters contribute to the relevant interaction estimates.

These results support a joint model--interface view of agent reliability.
Some operational distinctions must be preserved by the runtime regardless of
model capability, whereas other interface mechanisms overlap with behaviors a
stronger or more conservative model may already implement. The value of an
interface mechanism is therefore conditional on both the failure mode and the
agent policy that consumes it.

\subsection{Summary}

Across the three-model pooled analysis, all seven pre-specified primary
comparisons satisfy their planned inferential criteria after multiplicity
correction. The pattern is more informative than a single aggregate success
score. Selective discovery changes the information cost of capability
selection; resumable invocation and durable state address distinct execution
failures; effect semantics stabilize external actions under uncertain
commitment, stale state, and permission drift; and verification corrects
terminal beliefs that disagree with the resulting world state.

Taken together, these results support the central distinction developed in
Section~\ref{sec:model}: a tool may be callable while still hiding or failing
to stabilize distinctions that matter for the agent's next action. Interface
semantics can reduce that operational ambiguity either by exposing
authoritative state or by making continuation safe across multiple possible
states. Model capability changes the magnitude of some of these effects, but
does not eliminate the role of the interface itself.

\section{Limitations and Future Work}

Our experiments are designed for \emph{mechanism identification} rather than
for estimating how often the corresponding failures occur in deployed agent
systems. The workloads intentionally activate conditions such as transient
interruption, process-state loss, post-commit response loss, stale state,
permission drift, and incorrect terminal reports. The resulting effect sizes
should therefore be interpreted as conditional evidence about what an
interface mechanism does when the relevant failure occurs, not as estimates
of real-world failure prevalence.

The evaluation also trades breadth for experimental control. Most mechanism
experiments use resettable synthetic environments, with selected effect-safety
findings replicated on a persistent SQLite backend. This persistent-backend
replication reduces dependence on an in-memory simulator, but it does not
establish that the same effect sizes will hold across payment systems,
workflow engines, cloud services, enterprise APIs, or other production
backends. Extending the controlled design to additional independently
implemented systems is an important direction for future work.

Our adaptive evaluation covers three contemporary LLM families. This is
sufficient to reveal meaningful model--interface heterogeneity, but not to
establish model-independent universality. In particular, recovery mechanisms
are stable across the evaluated models, whereas the marginal value of
verification varies substantially with model policy. Future work could study
how interface value changes with model capability, controller design, and
agent conservatism, and whether some interface mechanisms act as complements
to model capability while others partially substitute for it.

The present study also focuses on reliability, recovery, context exposure, and
external-effect safety rather than production-scale performance. We record
local stage-level timing, but do not evaluate network queueing, geographically
distributed execution, or high-concurrency service behavior. Likewise,
observable execution and minimal structured outputs are part of the broader
\AFT model but receive weaker independent efficacy evidence than discovery,
resume, durable state, effect semantics, and verification. Dedicated
event-loss, parser-repair, and large-scale runtime experiments remain natural
extensions.

Finally, operational semantics are not always recoverable automatically from
legacy systems. Properties such as idempotency scope, authority, compensation,
or the evidence required to establish a postcondition may require explicit
developer or policy input. Future work should therefore consider not only
richer interface contracts, but also methods for deriving, validating, and
governing those contracts safely---including stronger trust models for tool
metadata, finer-grained authority and approval semantics, and integration with
production observability and audit systems.

\section{Conclusion}

A syntactically valid tool call is not sufficient for reliable autonomous
operation. Once execution becomes stateful or effectful, an agent must also be
able to determine what happened, which distinctions remain relevant, and what
continuation is safe. If two operational states require different actions but
appear identical through the available interface, additional model reasoning
alone cannot guarantee the correct decision.

This paper studies that gap between \emph{callability} and
\emph{operability}. We formulate \AFT as a set of tool-interface semantics for
capability discovery, execution lifecycle and recovery, external effects,
machine-readable results, and postcondition verification, and use \AFTBench
to isolate their effects while holding the task, backend, failure, controller,
and model fixed.

Across controlled deterministic experiments, a persistent backend, and three
adaptive LLM families, the results show a consistent mechanism-specific
pattern. Selective discovery sharply reduces context exposure without
sacrificing capability recall; resumable invocation and durable execution
state resolve distinct recovery failures; explicit effect semantics reduce
duplicate and unsafe external actions; and verification corrects terminal
claims that disagree with the resulting world state. The magnitude of some
effects depends on the model, but the broader conclusion does not reduce to a
model-capability comparison.

The central lesson is therefore not simply that better tools improve agents.
Reliable autonomous tool use depends on whether the interface preserves or
stabilizes the \emph{action-relevant distinctions} needed to choose what to do
next. A tool may be callable while remaining operationally ambiguous. Building
more capable agents will therefore require not only stronger models, but
interfaces whose semantics make safe continuation possible when execution and
external state become uncertain.

\clearpage
\bibliographystyle{unsrtnat}
\bibliography{references}

\end{document}